\documentstyle[12pt,epsfig]{article}
\makeatletter
\ifx\c@tmpcounter\UNDEFINED%
	\newcounter{tmpcounter}\else\fi  
\ifx\tmplength\UNDEFINED%
	\newlength{\tmplength}\else\fi   
\ifx\tmpbox\UNDEFINED%
	\newsavebox{\tmpbox}\else\fi    
\makeatother

\newcommand{\mbold}[1]{\mbox{\boldmath $ #1 $}}
\newcommand{\be}{\begin{equation}}
\newcommand{\ee}{\end{equation}}
\newcommand{\benn}{\begin{displaymath}}		
\newcommand{\eenn}{\end{displaymath}}		
\newcommand{\ba}{\begin{eqnarray}}	
\newcommand{\ea}{\end{eqnarray}}
\newcommand{\bann}{\begin{eqnarray*}}	
\newcommand{\eann}{\end{eqnarray*}}		

\def\Romannumber#1{\expandafter\uppercase\expandafter{\romannumeral #1}}
\def\LRdescriptionOPT{}	
\makeatletter%
\newenvironment{LRdescription}[1]%
{
  \begin{list}{#1}{
	\settowidth{\labelwidth}{#1}
    \setlength{\leftmargin}{\labelwidth}
	\setlength{\labelsep}{0pt}
	
\LRdescriptionOPT}%
\@topsepadd=0pt\@topsep=0pt%
}{\end{list}%
 }  
\def\lrdescriptionOPT{}	
{
  \begin{list}{#1}{
	\settowidth{\labelwidth}{#1}
    \setlength{\leftmargin}{\labelwidth}
	\setlength{\labelsep}{0pt}
	
\lrdescriptionOPT}%
}{\end{list}%
 }  
\makeatother%
\newcounter{hours}\newcounter{minutes}
\newcommand{\Time}{\setcounter{tmpcounter}{\number\time}%
      \setcounter{hours}{\value{tmpcounter}}%
      \setcounter{minutes}{\value{tmpcounter}}%
      \global \divide \value{hours} by 60\relax%
      \setcounter{tmpcounter}{\value{hours}}%
      \global \multiply \value{tmpcounter} by 60\relax%
      \global \advance \value{minutes} by -\value{tmpcounter}\relax%
      \ifnum\value{hours}<10 0\fi\thehours:%
      \ifnum\value{minutes}<10 0\fi\theminutes}
\def\todayEU{\number\day \  \ifcase\month\or
  January\or February\or March\or April\or May\or June\or
  July\or August\or September\or October\or November\or December\fi
   \  \number\year}

\newcommand{\lap}%
{\raisebox{-0.5ex}{$\stackrel{\scriptstyle <}{\scriptstyle \sim}$}}
\newcommand{\gap}%
{\raisebox{-0.5ex}{$\stackrel{\scriptstyle >}{\scriptstyle \sim}$}}
\def\opm{%
{\cal\rlap{$\scriptstyle\bigcirc$}%
\raisebox{0.05em}{$\scriptscriptstyle\kern0.1em\pm$}}}

\makeatletter
\def\eqsecnumbering{%
	\let\section@ori=\section
	\def\theequation{{\bf\thesection.}\arabic{equation}}
	\def\section{\setcounter{equation}{0}\section@ori}
}
\makeatother
\makeatletter
\newlength{\lr@listTW}
\newlength{\lr@listLW}
\def\lr@list#1{\def\lr@listDL{#1}%
\settowidth{\lr@listLW}{\lr@listDL}%
\lr@listTW=\linewidth%
\addtolength{\lr@listTW}{-\lr@listLW}%
\hbox to\lr@listLW{}\begin{minipage}[t]{\lr@listTW}
}
\def\lrlist{%
\def\lr@listPAR{\noindent}%
\def\lr@listPL[##1] {%
\lr@listPAR\raisebox{0pt}[0pt][0pt]{\llap{%
\parbox[t]{\lr@listLW}{##1}}}%
\def\lr@listPAR{\par\noindent}}%
\def\item{\@ifnextchar [{\lr@listPL}{\lr@listPDL}}%
\def\lr@listPDL{%
\lr@listPAR\raisebox{0pt}[0pt][0pt]{\llap{%
\parbox[t]{\lr@listLW}{\lr@listDL}}}%
\def\lr@listPAR{\par\noindent}}%
\lr@list}%
\def\endlrlist{\end{minipage}}
\makeatother
\def\figurename{Fig.}    
\ifx\LRcaptionwidth\UNDEFINED
\newlength{\LRcaptionwidth}\else\fi 
\LRcaptionwidth=\textwidth
\begin{document}
\hbadness=10000
\ifx\hyperref\undefined\else\errmessage{HYPER DISABLED}\fi
\setcounter{page}{1}
\thispagestyle{empty}
\begin{samepage}
\title{%
{\normalsize\sf%
{\sl August 1995\/}\hfill 
\begin{tabular}[t]{l}
GSI-Preprint-95-46 \\
hep-ph/9508318
\end{tabular}}\\*[5ex]
{\bf Photon Intensity Interferometry\\ for Expanding Sources}
}
\author{Leonid V. Razumov%
\thanks{E.Mail: \sf\,\, razumov@clri6a.gsi.de\ \ 
(URL$\Rightarrow$ http://www.gsi.de/groups/the/razumov.html)}\,\,, 
\quad Hans Feldmeier\setcounter{footnote}{3}%
\thanks{E.Mail: \sf\,\, h.feldmeier@gsi.de}
}

\date{Gesellschaft f\"ur Schwerionenforschung, Postfach 110552,\\
 D-64220 Darmstadt, Germany}
\maketitle
\setcounter{footnote}{0}
\thispagestyle{empty}
\begin{center}
{\large\bf Abstract}\\[2ex]
\end{center}
{\narrower\narrower\par\noindent \small
Using Quantum Field Theory we derive a general formula for the
double inclusive spectra of photons radiated by a system in local
equilibrium. The derived expression differs significantly from the one
mostly used up to now in photon intensity interferometry of heavy--ion
collisions.  We present a covariant expression for double inclusive spectra
adapted for usage in numerical simulations. Application to a schematic
model with a Bj\o rken type expansion gives strong evidence for the need
of reinvestigating photon--photon correlations for expanding sources.
\par}
\vskip 15mm
\noindent
\centerline{\large\sl Submitted to Physics Letters B\/.}
\end{samepage}
\newpage
\noindent
Double inclusive spectra of hard photons radiated at the early stage of
heavy--ion collisions have become an important subject of theoretical
\cite{Neuh86,SriKap93a,TimPlumRazWein94} and experimental
\cite{GANIL94} investigations.
Since these photons originate from very energetic collisions and interact
very little with the matter they are traversing, they may carry
signatures of the hot and dense phase of the matter, in particular of
the quark--gluon plasma.

Except for the work of D.~Neuhauser \cite{Neuh86} in which he treated
a static source, usually one is analyzing the correlations in two--photon
coincidence measurements with expressions based on the corresponding
ones for scalar bosons \cite{Pratt84} augmented with a degeneracy
factor which is supposed to take care of the spin structure of the
photons. 

The aim of this work is to derive in a quantum--field treatment 
covariant expressions for the two--photon density and herewith the
corresponding correlation function. In particular we pay attention to
charge conservation and to the fact that massless photons have only two
helicity eigenstates. Due to these properties averaging over all spin
directions results in a reduction factor of one--half compared to the
scalar case only if the momenta of the two photons are parallel. For a
finite angle, however, the interference is reduced, so that the deduced
size of the source is smaller than the one obtained from the
correlation function for scalar bosons. 
For Bj\o{}rken type hydrodynamics
we estimate a difference in the apparent source size of about 30\% in
rapidity. Therefore, we conclude that the ``pion--inspired'' formulas for
the Bose--Einstein correlation function is not applicable for
photon intensity interferometry studies, especially if one deals with an
expanding source. 

The finally measured number of photons per invariant momentum bin with
polarization $\,\lambda\,$ is given by the density matrix of the final
state $\rho_f$ which can be expressed in terms of the initial density
matrix $\rho_i$ and the $S$--matrix using the relation 
\mbox{$\rho_f = S \rho_i S^{\dag}$}.
\be \label{P1:def} 
P_1^{\lambda}(\mbold{k})\ \equiv\ k^0 \frac{d N^{\lambda}}{d^3\mbold{k}}
	= Tr\{ \rho_f c_{\lambda}^{\dag}(\mbold{k}) 
	  c_{\lambda}(\mbold{k})\}
	= Tr\{ \rho_i S^{\dag} c_{\lambda}^{\dag}(\mbold{k}) 
	  c_{\lambda}(\mbold{k}) S \}
\ee
The same consideration as above leads to the expression for the  number 
of photon pairs 
\ba \label{P2:def}
P_2^{\lambda_1\;\lambda_2}(\mbold{k}_1,\mbold{k}_2)\
	&\equiv&\ k^0_1k^0_2 \frac{d^2 N^{\lambda_1 \lambda_2}}%
	{d^3 \mbold{k}_1 d^3 \mbold{k}_2}
= Tr\{\rho_f c_{\lambda_2}^{\dag}(\mbold{k}_2) 
	c_{\lambda_1}^{\dag}(\mbold{k}_1)
	c_{\lambda_1}(\mbold{k}_1)
	c_{\lambda_2}(\mbold{k}_2)\}\ = \nonumber\\
&=&\ Tr\{\rho_iS^{\dag} c_{\lambda_2}^{\dag}(\mbold{k}_2) 
	c_{\lambda_1}^{\dag}(\mbold{k}_1)
	c_{\lambda_1}(\mbold{k}_1)
	c_{\lambda_2}(\mbold{k}_2) S\}\;\ . \quad
\ea
If we separate the interaction part of the Lagrangian into 
$L^{int}_s(x)$ for strong interactions and 
$J^{\mu}(x)A_{\mu}(x)$ for the coupling to the electromagnetic field the
\mbox{$S$--matix} can be written in the interaction picture  by the
time--ordered exponential \cite{BjorkenDrell65}
\be \label {S:def}
S\ =\ {\cal T}\exp\left\{i\int d^4x
\Big( L^{int}_s(x) + J^{\mu}(x)A_{\mu}(x)\Big) \right\}\;\ .\quad
\ee

To proceed further we assume \cite{Machl89} that the initial state does
not contain hard photons, which means
\be \label{rhoi:def}
c_{\lambda}(\mbold{k})\rho_i\ =\ 
\rho_i c^{\dag}_{\lambda}(\mbold{k}) \ =\ 0 \;\ .\quad
\ee
The condition (\ref{rhoi:def}) together with the identities
\ba 
\Big[ c_{\lambda}(\mbold{k})\,,\, S \Big] \ &=&\ 
i {\cal T} \Big(J_{\lambda}(k) S \Big) \quad, 
					\label{commut:cS}	\\\relax
\Big[ S^{\dag}\,,\, c^{\dag}_{\lambda}(\mbold{k}) \Big] \  &=&\  
-i \tilde{\cal T} \Big(S^{\dag} J^{\,\,\dag}_{\lambda}(k) \Big)
\quad, \nonumber
\ea
where $k^{\mu}=(|\mbold{k}|,\mbold{k})$ and
$J_{\lambda}(k)$ is the Fourier transform of the transverse current
operator given by
\ba
J_{\lambda}(k)\ &\equiv&\  
	\epsilon^{\!\!^{\scriptstyle *}\lambda}_{\mu}(k)
	\!\int\! d^4\!x\,{\rm e}^{ixk} J^{\mu}(x)
						\label{J_lambda:def} \\
J^{\,\,\dag}_{\lambda}(k)\ &\equiv&\  
	\epsilon^{\lambda}_{\mu}(k)
	\!\int\! d^4\!x\,{\rm e}^{-ixk} J^{\mu}(x) \quad, \nonumber
\ea
allow us to write one-- and two--body densities via the chronological 
(${\cal T}$) and antichronological ($\tilde{\cal T}$) products of the
current operators: 
%
\ba
P_1^{\lambda}(\mbold{k})\ &=&\ Tr \bigg\{ \rho_i %
	\tilde{\cal T} \Big( S^{\dag} J^{\,\,\dag}_{\lambda}(k) \Big)
	{\cal T}\Big( J_{\lambda}(k) S \Big) \bigg\}
							\label{P1:S} \\
P_2^{\lambda_1 \lambda_2}(\mbold{k}_1, \mbold{k}_2)\ &=&\
	Tr \bigg\{ \rho_i \tilde{\cal T} \Big( S^{\dag}%
	J^{\,\,\dag}_{\lambda_2}(k_2) J^{\,\,\dag}_{\lambda_1}(k_1) \Big)
	{\cal T} \Big( 
	J_{\lambda_1}(k_1) J_{\lambda_2}(k_2) S \Big)\bigg\} \quad .
							\label{P2:S}
\ea
All currents in the expressions written above are operators in the {\sl
interaction} representation. Utilizing properties of time
ordering and of the $S$--matrix we can write the one-- and two--photon
spectra in terms of the current operators in the {\sl
Heisenberg} representation ($\hat{J})$ \cite{RazWein95} as
\ba
P_1^{\lambda}(\mbold{k})\ &=&\ Tr \bigg\{ \rho_i %
	\hat{J}^{\,\,\dag}_{\lambda}(k) \hat{J}_{\lambda}(k) \bigg\}
							\label{P1:H} \\
P_2^{\lambda_1 \lambda_2}(\mbold{k}_1, \mbold{k}_2)\ &=&\ 
	Tr \bigg\{ \rho_i \tilde{\cal T} \Big( %
	\hat{J}^{\,\,\dag}_{\lambda_2}(k_2)
	\hat{J}^{\,\,\dag}_{\lambda_1}(k_1) \Big)
	{\cal T} \Big( 
	\hat{J}_{\lambda_1}(k_1) \hat{J}_{\lambda_2}(k_2) \Big)\bigg\}
				\quad.			\label{P2:H}
\ea
Up to now we have made no approximation to tackle the unsolvable
many--body problem. Expressions (\ref{P1:H}) and (\ref{P2:H}) are still
exact but they are recast into a form which is better suited for 
approximations than the original definitions (\ref{P1:def}) and
(\ref{P2:def}). Formally they contain only operators of the strongly
interacting system.

In the following we apply eqs. (\ref{P1:H}) and (\ref{P2:H}) to a highly
excited system which due to short--ranged hard collisions of the charged
constituents is radiating energetic photons. The collisions cause
locally rapid changes in the electric charge current density
$J^{\mu}(x)\,$, which result in non--zero Fourier components in
$J^{\mu}(k)\,$ for large $|\mbold{k}|\,$. The energetic photons are
weakly interacting so that, once they are created, they are assumed to
leave the hadron system without further interactions. Therefore, for
calculating the final photon distributions one needs a local production
rate which is then integrated over space and time. Eq. (\ref{P1:H}) is
of this type. In coordinate space it reads
\be \label{P1:xdx}
P^{\lambda}_1(\mbold{k})= 
\epsilon^{\!\!^{\scriptstyle *}\lambda}_{\mu_1}(k)
\epsilon^{\lambda}_{\mu_2}(k)\!\!
\int\!\!d^4\!\bar{x}\!\!\int\!\!d^4\!\Delta{}x\, {\rm e}^{-ik\Delta x}
\,Tr\Big\{\rho_i \hat{J}^{\mu_1}(\bar{x}+\frac{\Delta x}{2})
\hat{J}^{\mu_2}(\bar{x}-\frac{\Delta x}{2}) \Big\}
\ee
where the local rate is given by the integral over $\Delta x\,$ of the
current--current correlator. 

The following approximations, which will lead to an expression for the
correlator adapted for usage in numerical simulations for heavy--ion
collisions, are based upon three physical assumptions:
\begin{LRdescription}{$(iii)$\hbox{\,\ }}
\item[\hfill$(i)$\hbox{\,\ }]  The hadronic correlations in the
system are of short range in space--time. The typical correlation length
is about the mean free path $\lambda_s$ of the strongly interacting
particles. Therefore,
\benn <\!A(x) B(y)\!> \approx <\!A(x)\!> <\!B(y)\!> \qquad
\mbox{\rm if}\;\ |\mbold{x} - \mbold{y}| \;\gap\; \lambda_s \;\ 
\mbox{\rm or}\;\ |x^0 - y^0| \;\gap\; \lambda_s\ .  
\eenn
\item[\hfill$(ii)$\hbox{\,\ }] The hadronic mean free path
$\lambda_s$ is much less than the characteristic size of variations of
the macroscopic variables in space and time, denoted by $L\,$.
\item[\hfill$(iii)$\hbox{\,\ }] Only hard photons with high momenta
$|\mbold{k}|$ are considered
$\,(\,|\mbold{k}| \lambda_s \;\gap\; 1\,)\,$, for which
the slowly varying collective current
$<\!\!\hat{J}^{\mu}(x)\!\!>$ (proportional to the collective 4--velocity
field $u^{\mu}(x)\,$) does not contribute, i.e. 
$<\!\!\hat{J}^{\mu}(k)\!\!> \approx 0 $ for high enough $|\mbold{k}|\,$.
\end{LRdescription}
For instance, applying these assumptions to (\ref{P1:xdx}) means that
the integral in $\Delta x$ effectively spreads over the 4--volume of the
order $\lambda_s^4$ and the single inclusive spectrum $P_1(\mbold{k})\,$
is therefore proportional to $L^4 \lambda^4_s\;$.
 
Inside the trace on the right--hand side of (\ref{P1:xdx}) the mean time
$\bar{x}^0\,$ can be moved over to the statistical operator such that
\be \label{JJ:trick}
Tr\Big\{\rho_i \hat{J}^{\mu_1}(\bar{x}+\frac{\Delta x}{2})
\hat{J}^{\mu_2}(\bar{x}-\frac{\Delta x}{2}) \Big\} =
Tr\Big\{\rho(\bar{x}^0) \hat{J}^{\mu_1}(\tilde{x}+\frac{\Delta x}{2})
\hat{J}^{\mu_2}(\tilde{x}-\frac{\Delta x}{2}) \Big\}\ ,\qquad
\ee
where $\tilde{x}=(0,\mbold{x})$ and 
	$\rho(\bar{x}^0)={\rm e}^{-iH\bar{x}^0} 
	\rho_i {\rm e}^{iH\bar{x}^0}$ 
is the solution of the Liouville equation for the density matrix.  For
the case of local equilibrium, where the time--dependence enters the
density matrix $\,\rho(\bar{x}^0)\,$
only through the thermodynamic quantities, like for
example temperature $T(\bar{x}^0,\mbold{x})\,$, 4--velocity
$\,u^{\mu}(\bar{x}^0,\mbold{x})\,$ etc., the identity (\ref{JJ:trick})
is a starting point to express the current--current correlator in terms
of thermodynamic variables.

The double inclusive spectrum is given by the four--point correlator
(\ref{P2:H}) and like any other Green function can be written in terms
of connected parts as
\begin{samepage}
\ba 
P_2^{\lambda_1 \lambda_2}(\mbold{k}_1, \mbold{k}_2) =&&
	P_1^{\lambda_1}(\mbold{k}_1) P_1^{\lambda_2}(\mbold{k}_2) 
  + |\!\!<\!\!\hat{J}^{\dag}_{\lambda_1}(k_1)
	\hat{J}_{\lambda_2}(k_2)\!\!>\!\!|^2
  + |\!\!<\!\!{\cal T} \Big( \hat{J}_{\lambda_1}(k_1)
	\hat{J}_{\lambda_2}(k_2) \Big)\!\!>\!\!|^2 \qquad
						\nonumber \\
&& \;+\; \ll\!\! \tilde{\cal T} \Big(
	\hat{J}^{\,\,\dag}_{\lambda_2}(k_2)
	\hat{J}^{\,\,\dag}_{\lambda_1}(k_1) \Big)
	{\cal T} \Big( \hat{J}_{\lambda_1}(k_1)
	\hat{J}_{\lambda_2}(k_2) \Big) \!\!\gg\ .\qquad
\label{P2:JJ}
\ea
\end{samepage}
All terms in (\ref{P2:JJ}) involve four space--time integrations.  In
coordinate space a connected correlator disappears if the distance
between any pair of its space--time points exceeds the correlation
length. The connected four--point correlator $\,\ll\!\!\cdots\!\!\gg\,$
(last term on the right--hand side of (\ref{P2:JJ})\,) is only non--zero
if the relative distance of all its space--time points are within
$\,\lambda_s\,$.  Therefore, this term is proportional to $\,L^4
(\lambda^4_s)^3\,$ and hence, due to condition $\,(ii)\,$, is smaller by
factor of $\,(\lambda_s/L)^4\,$ when compared to all other terms which are
proportional to $\,(L^4)^2(\lambda^4_s)^2\,$.
The third term on the r.h.s of (\ref{P2:JJ}), 
	$\,|\!\!<\!\!{\cal T} 
	\Big( \hat{J}_{\lambda_1}(k_1)
	\hat{J}_{\lambda_2}(k_2) \Big)\!\!>\!\!|^2\,$,
contains an integral $\,\int\!d^4\!\bar{x}\,
	\exp(i(|\mbold{k}_1|+|\mbold{k}_2|)\bar{x}^0)\,\cdots \;$ 
in which the energies of the two photons add up in the phase
factor. This implies that only momenta which fulfill the condition
$\,(|\mbold{k}_1|+|\mbold{k}_2|)\cdot L \;\lap\; 1 \,$ contribute
appreciably to this term. 
These momenta are much smaller than required by
conditions $\,(ii)\,$ and $\,(iii)\,$ and hence we shall drop this term.

After these approximations based on the statistical properties of the
 emitting source the one-- and two--photon spectra (\ref{P1:H}) and
 (\ref{P2:H}) are given by the current--current correlator
 $\,<\!\!\hat{J}^{\mu_1 \dag}(k_1)\hat{J}^{\mu_2}(k_2)\!\!>\,$ only.
 Because the electrical current is conserved
 $(\partial_{\mu}\hat{J}^\mu(x)=0)$ the correlator
 must be transverse ``from both sides'':
\be \label{kJJ=0}
k_{1 \mu_1}<\!\!\hat{J}^{\mu_1 \dag}(k_1)\hat{J}^{\mu_2}(k_2)\!\!> =
<\!\!\hat{J}^{\mu_1 \dag}(k_1)\hat{J}^{\mu_2}(k_2)\!\!>k_{2 \mu_2} = 0
\;\ .\qquad\ee
 In analogy to (\ref{P1:xdx}) the correlator can be written as
\ba
	<\!\!\hat{J}^{\mu_1 \dag}(k_1)\hat{J}^{\mu_2}(k_2)\!\!> =
	\sum\limits_{n}&&{\rm e}^{-i(k_1-k_2)\bar{x}_n}\!\!
	\int\limits_{\Omega_n}\!\!d^4\!\bar{x}
	{\rm e}^{-i(k_1-k_2)(\bar{x}-\bar{x}_n)}\!\!\int\!\!d^4\!\Delta x
	{\rm e}^{-i\Delta x (k_1+k_2)/2}\, 
						\hbox{\qquad}\nonumber\\
	\times&&Tr\Big\{\rho_i \hat{J}^{\mu_1}(\bar{x}+\frac{\Delta x}{2})
\hat{J}^{\mu_2}(\bar{x}-\frac{\Delta x}{2})\Big\}\;\ ,
\qquad \label{JJ:xdx}
\ea
where the whole space--time region occupied by the source is divided 
into cells with 4--volume $\,\Omega_n\,$ located at mean--coordinates
$\,\bar{x}_n\,$.
Under the assumptions $\,(i)-(iii)\,$ the size of the space-time cells
should be chosen to be about $\lambda_s$ so that the radiation of hard
photons from different cells can be considered as entirely
independent processes. This implies that the currents are conserved for
each cell separately and that the contribution of the cells to the total
current correlator is additive. Therefore, we parametrize (\ref{JJ:xdx})
as
\be \label{JJ:W}
<\!\!\hat{J}^{\mu_1 \dag}(k_1)\hat{J}^{\mu_2}(k_2)\!\!>\  
\stackrel{ansatz}{=\!=\!=} \
\sum\limits_{n} \Omega_n{\rm e}^{-i\bar{x}_n(k_1 - k_2)}
Q^{\mu_1\mu_2}(k_1,k_2|\bar{x}_n) w(k_1,k_2|\bar{x}_n)\;\ ,\quad
\ee
where the tensor $Q^{\mu_1 \mu_2}$ carries the tensorial structure
of the current--current correlator and the function $w$ describes the
strength.
The $Q$--tensor is explicitly transverse
\be \label{kQ=0}
\,k_{1\mu_1}Q^{\mu_1\mu_2}(k_1,k_2|x) =
Q^{\mu_1\mu_2}(k_1,k_2|x)k_{2\mu_2}\nolinebreak=\nolinebreak{}0\, 
\ee
and, therefore, ensures current conservations for every cell.
The tensor $Q^{\mu_1\mu_2}$ and the function $w$ 
are hermitian in the sense that 
$Q^{\mu_1\mu_2}(k_1,k_2|x)^{\mbox{\boldmath$*$}}
=Q^{\mu_2\mu_1}(k_2,k_1|x)$
and $w(k_1,k_2|x)^{\mbox{\boldmath$*$}}
=w(k_2,k_1|x)$. It is convenient to normalize the $Q$--tensor by
$\,Q^{\mu}_{\mu}(k,k|x) \equiv -2\,$.
From now on we will write an integral instead of the sum in 
(\ref{JJ:W}).

Altogether single and double inclusive photon spectra (no polarization
measured) can be expressed with the help of (\ref{JJ:W}) as follows%
\footnote{We take the sum over photon polarisations by means of
	\benn
		\!\sum\limits^2_{(\lambda=1)}\!%
		\epsilon^{\!\!^{\scriptstyle *}\lambda}_{\mu}(k)%
		\epsilon^{\lambda}_{\nu}(k)
		= - g_{\mu\nu} - k_{\mu}k_{\nu}/(sk)^2 
		+ (k_{\mu}s_{\nu} + s_{\mu}k_{\nu})/(sk)
		\eenn
($\,s^{\mu}\,$ is some reference 4-vector $\,s^2=1\,$),
where due to the transversality condition (\ref{kQ=0}) only the first term
on the right--hand side $\,(-g_{\mu\nu})\,$ really contributes to
(\ref{P1:W}), (\ref{P2:W}). 
}: 
\ba
P_1(\mbold{k})\ &\equiv\ & \!\sum^2_{(\lambda=1)} P_1^{\lambda}(\mbold{k}) 
 	= 2 \!\int\! d^4\!x\, w(k,k|x)
								\label{P1:W} \\
P_2(\mbold{k}_1,\mbold{k}_2) &\equiv&
	\!\sum^2_{(\lambda_1,\lambda_2=1)}\!\! 
	P_2^{\lambda_1 \lambda_2}(\mbold{k}_1,\mbold{k}_2) =
	P_1(\mbold{k}_1) P_1(\mbold{k}_2)
								\label{P2:W} \\
&+& \!\int\!d^4\!x d^4\!y \, 
	{\rm e}^{-i(x-y)(k_1-k_2)} 
	   w(k_1,k_2|x) w(k_2,k_1|y)
	Q^{\mu_1\mu_2}(k_1,k_2|x) 
	Q_{\mu_2\mu_1}(k_2,k_1|y) 
\quad \nonumber
\ea

To find an explicit form of the $Q$--tensor we assume that the radiating
system is in local equilibrium, that is, the radiation from an elementary
cell is isotropic in its local rest frame and therefore
$Q^{i_1i_2}=\delta^{i_1i_2}$ (where $\,i_1,i_2=1,2,3$).
The other components of the $Q$--tensor are then uniquely determined by the
transversality condition (\ref{kQ=0}) as 
$\,Q^{00}=\mbold{k}_1\mbold{k}_2/k^0_1k^0_2\,$, 
$\,Q^{i_10}=k^{i_1}_2/k^0_2\,$,  $\,Q^{0i_2}=k^{i_2}_1/k^0_1\,$.
In local thermal equilibrium the only other 4--vector entering is the
4--velocity $u^{\mu}(x)$ of the emitting cell which is located at point
$x\,$. Therefore, the only hermitian and transversal $Q$--tensor which
can be constructed from $k^{\mu}_1\,,\; k_2^{\mu}\,\;$ and $u^{\mu}(x)\,$,
and which in the rest frame of a cell (\,$u^{\mu}=(1,\mbox{\bf 0})$\,) 
reproduces the isotropic form given above, reads as follows:
\be \label{Q:leq}
Q^{\mu_1\mu_2}(k_1,k_2|x) = -g^{\mu_1\mu_2}  
- u^{\mu_1}(x)u^{\mu_2}(x)\frac{(k_1k_2)}{(k_1u(x))(k_2u(x))}
+ \frac{u^{\mu_1}(x)k_1^{\mu_2}}{(u(x)k_1)} 
+ \frac{k_2^{\mu_1}u^{\mu_2}(x)}{(u(x)k_2)}\ .
\ee
In eq.~(\ref{Q:leq}) we do not include  the term proportional to
$\,k_1^{\mu_1}k_2^{\mu_2}\,$ because it does not contribute to
observables and can be removed by an appropriate gauge transformation.

\begin{figure}
\begin{center}
\centerline{\epsfig{file=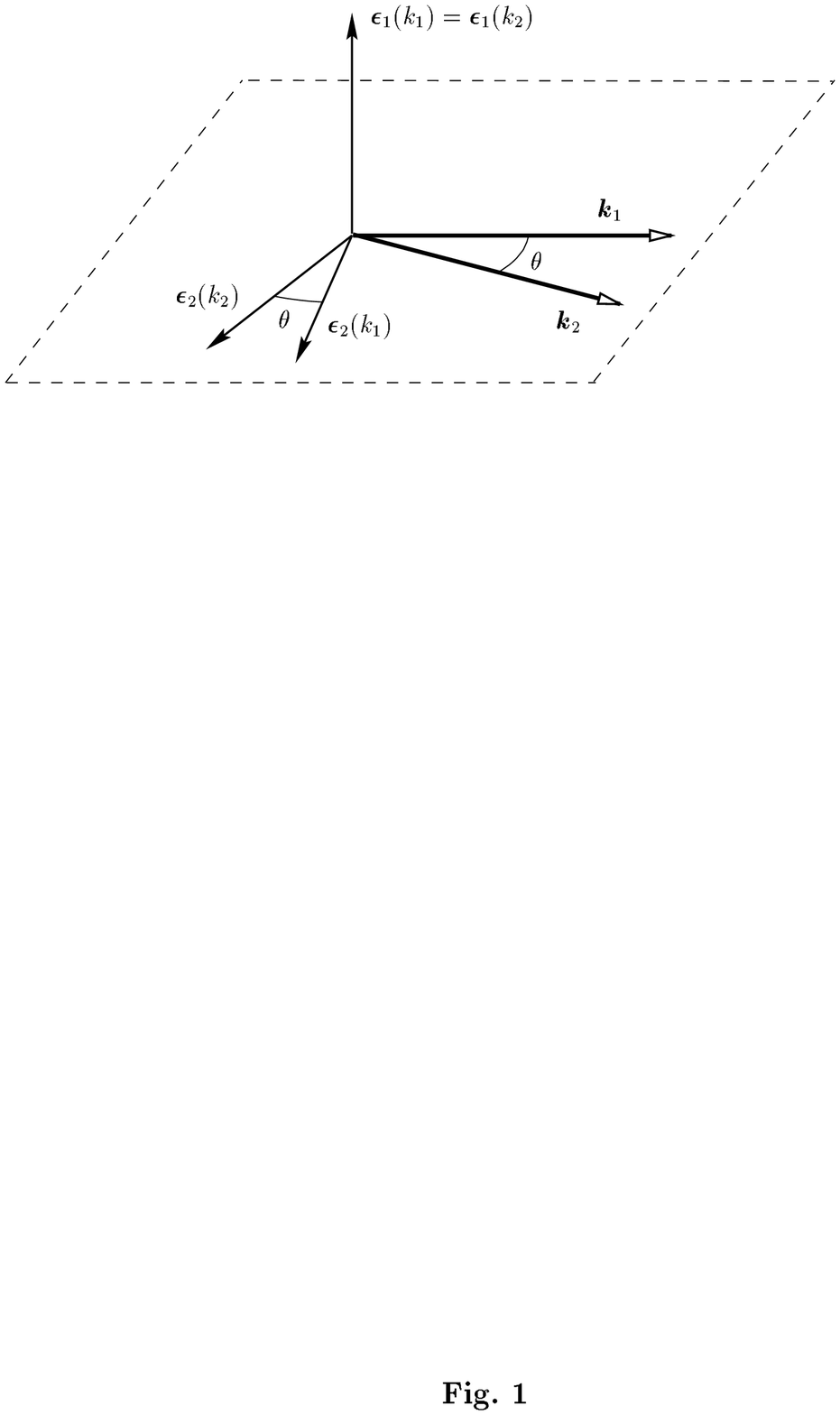,%
bbllx=70bp,bblly=580bp,bburx=465bp,bbury=766bp,clip=}}
\vskip 3mm
\begin{minipage}{\LRcaptionwidth}
{\refstepcounter{figure}\label{fig:pol}}
\sbox{\tmpbox}{\bf\figurename~\thefigure:\hbox{\ }}
\begin{LRdescription}{\usebox{\tmpbox}}
\item \footnotesize 
	The linear polarization vectors $\mbold{\epsilon}^{\lambda}(k)$ 
	for two photons with momenta 
	$\mbold{k}_1\,$ and $\mbold{k}_2\,$.
\end{LRdescription}
\end{minipage}
\end{center}
\end{figure}

After defining the $Q$--tensor all dynamic information on the photon
production is contained in the function $w(k_1,k_2|x)$. Inserting
(\ref{Q:leq}) into (\ref{P2:W}) one gets
\be 	\label{P2:leq}
\!\!\!%
P_2(\mbold{k}_1,\mbold{k}_2) \!=\! P_1(\mbold{k}_1) P_1(\mbold{k}_2) 
\!+\!\!\int\!d^4\!x d^4\!y \cos(\Delta{}x\Delta{}k)
	   R(k_1,k_2|x,y) w(k_1,k_2|x) w(k_2,k_1|y)\ ,
\ee
where we introduce the abbreviation 
\ba
&&R(k_1,k_2|x,y) \equiv
	Q^{\mu_1 \mu_2}(k_1,k_2|x)Q^{\mu_2 \mu_1}(k_2,k_1|y) = 
	\frac{\big(u(x)k_1\big)\big(u(y)k_2\big) -
	2\big(u(x)u(y)\big)\big(k_1k_2\big)}{%
\big(u(x)k_2\big)\big(u(y)k_1\big)} 
							\qquad	\nonumber \\
&&+\ \frac{\big(k_1k_2\big)}{\big(k_1u(x)\big)\big(u(x)k_2\big)} \,+\, 
\frac{1}{2}\frac{\big(u(x)u(y)\big)^2\big(k_1k_2\big)^2}{%
\big(k_1u(x)\big)\big(u(x)k_2\big)\big(k_1u(y)\big)\big(u(y)k_2\big)} 
\ +\ \Big(x \Leftrightarrow y \Big)\;\ .
			\quad \label{R:leq}
\ea
The $R$--function reflects the fact that photons are massless spin--1
particles and in this point expression (\ref{P2:leq}) differs
significantly from the one obtained in \cite{Pratt84} for pions.  The
latter has been used for photons of one single polarization in
\cite{SriKap93a,TimPlumRazWein94,SriKap93,SriKap94}.  Since the
experimental measurements will involve an averaging over polarizations,
a comparison with data requires the new result (\ref{P2:leq}).
Being very sensitive to the 4--velocity field $\,u^{\mu}(x)\,$ the
$R$--function can change drastically the interference of two photons for
a relativistically expanding source. But even for a static source the
polarization properties of photons are very important.
To illustrate this let us consider two photons with momentum 
$\,\mbold{k}_1\,$ and $\,\mbold{k}_2\,$ which are emitted from two cells
having the same 4--velocity $\,u(x)=u(y)=(1,\mbold{0})\,$. In that
case \cite{Neuh86} the $R$--function reduces to 
\be
R(k_1,k_2|x,y) = 1 + (\cos\theta)^2 = 
\sum\limits^2_{(\lambda_1,\lambda_2 = 1)}\Big(
\mbold{\epsilon}^{\lambda_1}(k_1)
\mbold{\epsilon}^{\lambda_2}(k_2)
\Big)^2\;\ ,
\qquad \label{R:ux=uy}
\ee
where
$\cos\theta=\mbold{k}_1\mbold{k}_2/(|\mbold{k}_1||\mbold{k}_2|)\,$. 
The physical reason is easily understood for this special case when one
considers the last part of equation (\ref{R:ux=uy}). The summation over
polarizations (here we choose the Coulomb gauge: 
$\,\epsilon^{\lambda}(k)=(0,\mbold{\epsilon}^{\lambda}(k))\,$ and 
$\,\mbold{k}\mbold{\epsilon}^{\lambda}(k)=0\,$) does not just lead to a
factor of two as implicitly assumed in
\cite{SriKap93a,TimPlumRazWein94,SriKap93,SriKap94}.
As illustrated by \figurename~\ref{fig:pol} only one direction of the
linear polarization can be chosen equal for both photons, whereas the
other polarization directions differ by the angle $\,\theta\,$ between
the two momenta. Therefore, the polarization overlap involves
$\,(\cos\theta)^2\,$ and is less than 2 for $\,\theta \neq 0\,$ (see
(\ref{R:ux=uy})). In a realistic situation the radiation comes of
course from many cells with different four--velocities and the full
structure of the $Q$--tensor has to be employed. Anyhow, for a
non--vanishing angle between the observed photon momenta the correlation 
is reduced compared to the ``pion inspired'' recipe
(\ref{P2:wrong}).

For application in dynamical models we recommend to use the following
formula: 
\be 	\label{P2:JmuJnu}
P_2(\mbold{k}_1, \mbold{k}_2) =
	P_1(\mbold{k}_1) P_1(\mbold{k}_2) 
  + \!<\!\!\hat{J}^{\mu_1 \dag}(k_1) \hat{J}^{\mu_2}(k_2)\!\!>
	<\!\!\hat{J}^{\dag}_{\mu_2}(k_2)
	\hat{J}_{\mu_1}(k_1)\!\!>\! 
\ee
and calculate the current--current correlator with (\ref{JJ:W}) keeping
the explicit form of the $Q$--tensor as given in (\ref{Q:leq}).  This is
equivalent to (\ref{P2:leq}) but involves only one $\,d^4\!x\,$
integration instead of $\,d^4\!x\,d^4\!y\,$ in (\ref{P2:leq}).

The dependence of $\,w(k_1,k_2|x)\,$ on $\,\Delta k = k_1 -k_2\,$ and 
$\,\bar{k}=(k_1 + k_2)/2\,$ has to be derived from a microscopic model
for the photon production. The only general statement which can be made
at this point is that due to condition $\,(i)\,$ the current--current
correlator decays in the local restframe as function of $\,\Delta k\,$
on the long scale $\,\lambda_s^{-1}\,$. In this paper we do not
consider a specific microscopic model for $\,w(k_1,k_2|x)\,$ but rather
investigate the role of the tensorial structure of the correlator on
photon--photon interferometry. Therefore we write 
$\,w(k_1,k_2|x) = w(\bar{k},\bar{k}|x) + 
{\cal O}(\Delta k^{\mu}\Delta k^{\nu})\;$. For simplicity we drop terms
of second order in $\,\Delta k\,$. 
The strength of photon production $\,w(\bar{k},\bar{k}|x)\,$ can to some
extend be estimated from the single inclusive cross section
 \cite{KapLicSeib92}.


Based on the
discussion above we propose to reexamine the photon intensity
interferometry in the spirit of 
\cite{SriKap93a,TimPlumRazWein94,SriKap93,SriKap94} but using 
(\ref{P2:JmuJnu}) for the double inclusive cross section.

Unfortunately until now the Bose--Einstein correlations of photons have
been studied using the formula for double inclusive spectra where the
photons are assumed to be scalar massless particles, but normalizing the
correlation function to $3/2$ instead of $2$
\cite{SriKap93,SriKap93a,SriKap94}.
That formula is just expression 
(\ref{P2:leq}) with $R(k_1,k_2|x,y) = 2\;$ and reads
\be 	\label{P2:wrong}
P_2^{wrong}(\mbold{k}_1,\mbold{k}_2) = P_1(\mbold{k}_1) P_1(\mbold{k}_2) 
+ 2 \!\int\!d^4\!x d^4\!y \, \cos(\Delta{}x\Delta{}k)
w(\bar{k},\bar{k}|x)w(\bar{k},\bar{k}|y)\;\ .  \qquad
\ee

\begin{figure}
\begin{center}
\centerline{\epsfig{file=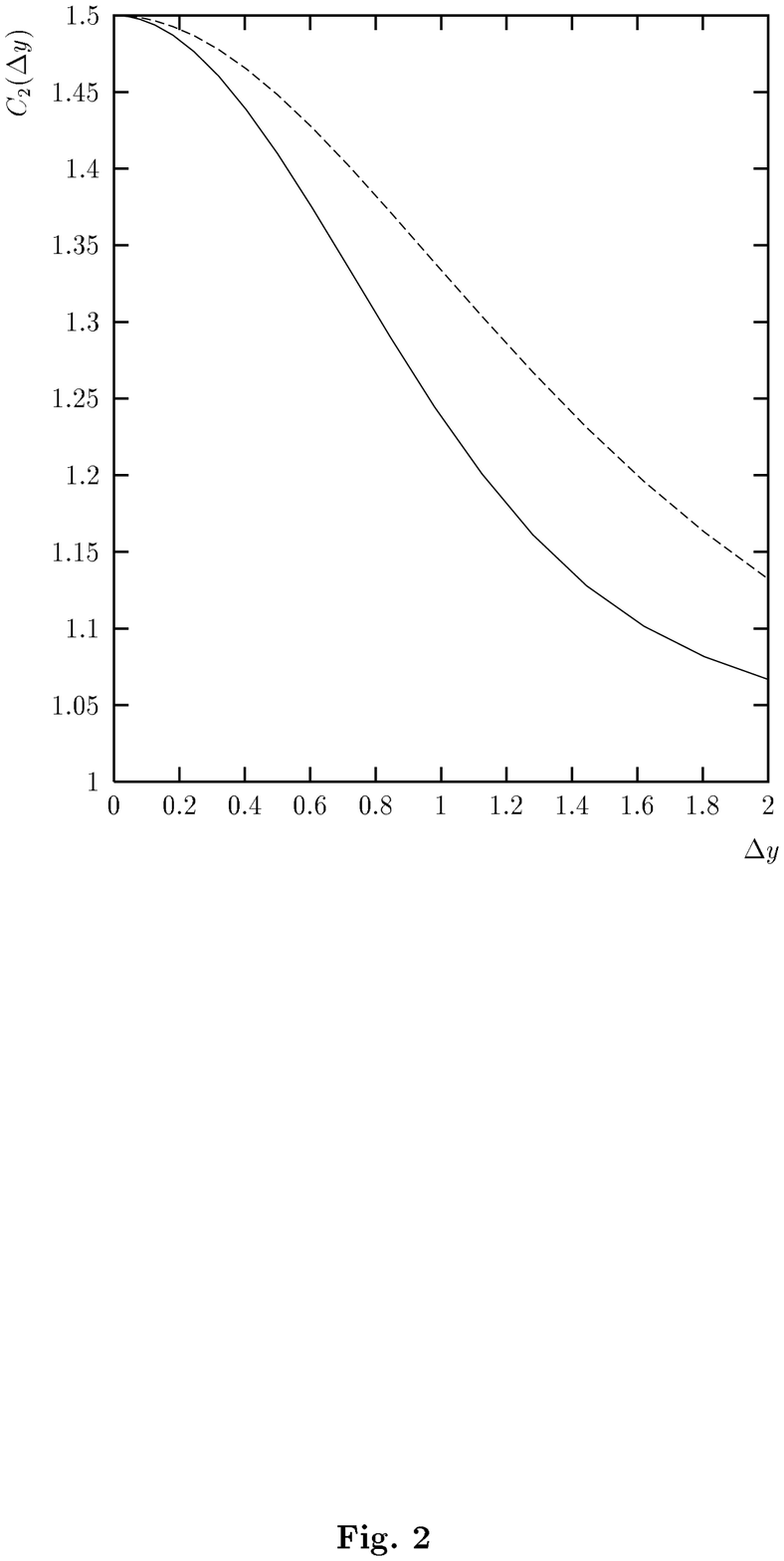,%
bbllx=140bp,bblly=390bp,bburx=500bp,bbury=780bp,clip=}}
\vskip 3mm
\begin{minipage}{\LRcaptionwidth}
{\refstepcounter{figure}\label{fig:hyd}}
\sbox{\tmpbox}{\bf\figurename~\thefigure:\hbox{\ }}
\begin{LRdescription}{\usebox{\tmpbox}}
\item \footnotesize 
Bose--Einstein correlation plot as a function of rapidity
difference $\Delta y$ at fixed transverse momenta
$\mbox{\bf k}_1^{\perp}=\mbox{\bf k}_2^{\perp}=100\; \mbox{\rm MeV} $ 
for Bj\o rken hydrodynamics with 
$T_i=200\; \mbox{\rm MeV}$, $T_f=140\; \mbox{\rm MeV}$ and initial 
proper--time $\tau_i=0.3\; fm/c\;$.\\
Solid line corresponds to the photon--photon correlations
(\ref{P2:JmuJnu}) (or equivalently (\ref{P2:leq})). \\ 
Dashed line represents the wrong ``pion--inspired''
expression (\ref{P2:wrong}).
\end{LRdescription}
\end{minipage}
\end{center}
\end{figure}

In order to demonstrate the size of the effect which arises from the
vector nature of the photon and the conservation of electric charge we
have performed numerical studies of photon intensity interferometry in
Bj\o rken hydrodynamics (equation of state $p = \epsilon/3$) using both
our formula (\ref{P2:JmuJnu}) (or equivalently (\ref{P2:leq})) and the
wrong one (\ref{P2:wrong}).  We assume that the photons are produced
from the expanding source in local equilibrium and parameterize the
photon production rate as
$\,w(k_1,k_2|x) \cong w(\bar{k},\bar{k}|x) = N\cdot (T(x))^2
\exp(-(\bar{k}u(x))/T(x))\,$ (for a more precise expression
cf. \cite{KapLicSeib92}).  The plot of the correlation function defined
as $C_2(k_1,k_2) \equiv P_2(k_1,k_2)/P_1(k_1)P_1(k_2) $ shows a
significant difference in the results based on the different formulas
(see \figurename~\ref{fig:hyd}). The correct expression gives less
correlations for non--zero relative rapidities which means that the
source size deduced from experimental data is overestimated by the wrong
formula (\ref{P2:wrong}).

In summary we should like to stress that the derivation of a basic
equation for the double inclusive spectrum is significantly modified by the
fact that photons are massless particles with spin 1 and that they are
produced by a conserved electric current.  Under the conditions
$\,(i)-(iii)\,$ which are fulfilled for high--energy photons in
relativistic heavy--ion collisions, the expression (\ref{P2:JmuJnu}) is 
a suitable starting point for photon intensity interferometry studies.

We would like to acknowledge the fruitful discussions with G.~Bertsch
and J.~Knoll as well as the comments of D.~Seibert. Especially we are
grateful to R.~Weiner and M.~Pl\"umer for critical remarks.

\newpage
\def\noopsort#1{} \def\SL#1{#1}

\end{document}